\def\ttitle{Deep Learning for Absorption-Image Analysis}
\def\kkeywords{
  machine learning,
  imaging
}

\documentclass[pra,reprint,superscriptaddress,showkeys]{revtex4-2}
\usepackage[letterpaper, margin=.75in]{geometry}
\usepackage{graphicx} 
\usepackage{svg}
\usepackage{multirow}%
\usepackage{amsmath,amssymb,amsfonts}%
\usepackage{amsthm}%
\usepackage{mathrsfs}%
\usepackage[title]{appendix}%
\usepackage{xcolor}%
\usepackage{textcomp}%
\usepackage{booktabs}%
\usepackage{algorithm}%
\usepackage{algorithmicx}%
\usepackage{algpseudocode}%
\usepackage{listings}%
\usepackage{epstopdf}
\usepackage{tabularx}
\usepackage[pdfpagelabels=true]{hyperref}
\usepackage{hypernat}
\usepackage[nolist]{acronym}
\usepackage{fancyhdr}
\usepackage{./sty/shortcuts} 
\usepackage{xr}

\setboolean{showcomment}{true}

\usepackage{lineno}

\usepackage{eso-pic} 
\makeatletter
  \AddToShipoutPicture{%
    \setlength{\@tempdimb}{0.2in}%
    \setlength{\@tempdimc}{.5\paperheight}%
    \setlength{\unitlength}{1pt}%
    \put(\strip@pt\@tempdimb,\strip@pt\@tempdimc){%
      \makebox(100,5)[l]{\rotatebox{90}{%
        \textcolor[gray]{.65}{%
          \fontsize{1cm}{1cm}%
        }%
      }%
    }%
  }%
}
\makeatother

\hypersetup{
    naturalnames=true,
    colorlinks=true,
    linkcolor=blue,
    pdfpagemode=UseNone,
    pdfstartview=FitH,
    pdftitle={\ttitle},
    pdfauthor=Air Force Research Laboratory,
    pdfsubject={Machine learning for experimental atomic physics},
    pdfkeywords={\kkeywords},
}

\newcolumntype{L}[1]{>{\raggedright\let\newline\\\arraybackslash}m{#1}}
\newcolumntype{C}[1]{>{\centering\let\newline\\\arraybackslash}m{#1}}
\newcolumntype{R}[1]{>{\raggedleft\let\newline\\\arraybackslash}m{#1}}

\newcommand{\distA}[1]{%
  Approved for public release; distribution is unlimited.  Public Affairs %
  release approval %
  #1.
}

\begin{document}
\title{\ttitle}

\author{Jacob Morrey}
\thanks{These two authors contributed equally to this work}
\author{Isaac Peterson}
\thanks{These two authors contributed equally to this work}
\affiliation{%
  Universities Space Research Association, %
  %
}

\author{Robert H. \surname{Leonard}}
\author{Joshua M. \surname{Wilson}}
\author{Francisco \surname{Fonta}}

\affiliation{Space Dynamics Laboratory, Quantum Sensing \& Timing, North Logan, UT 84341, USA}

\author{Matthew B. \surname{Squires}}
\affiliation{\AFRLAddress}

\author{Spencer E. \surname{Olson}}
\def\AFRLAddress{%
  Space Vehicles Directorate, %
  Air Force Research Laboratory, %
  3550 Aberdeen Ave SE, %
  Kirtland Air Force Base, %
  87117, %
  New Mexico, %
  USA%
}
\affiliation{\AFRLAddress}

\date{\today}

\begin{abstract}
      The quantum state of ultracold atoms is often determined through
      measurement of the spatial distribution of the atom cloud.  Absorption
      imaging of the cloud is regularly used to extract this spatial information.
      Accurate determination of the parameters which describe the spatial
      distribution of the cloud is crucial to the success of many
      ultracold atom applications.  In this work,  we present modified
      deep learning image classification models for image regression.
      To overcome challenges in data collection, we train the model on
      simulated absorption images.
      We compare the performance of the deep learning models to least-squares
      techniques and show that the deep learning models achieve accuracy
      similar to least-squares, while consuming significantly less
      computation time.  We compare the performance of models which
      take a single atom image against models which use an atom image plus other
      images that contain background information, and find that both models
      achieved similar accuracy.
      The use of single image models will enable single-exposure absorption
      imaging, which simplifies experiment design and eases imaging
      hardware requirements.
      The code used to train and evaluate these models is open-source.
\end{abstract}

\keywords{\kkeywords}

\maketitle \thispagestyle{fancy}

\DeclareRobustCommand{\ThreeDLS}{$3\times1$-dimensional least-squares}
\begin{acronym}
  \acro{ML}{machine learning}
  \acro{PCA}{principal component analysis}
  \acro{OD}{optical density}
  \acro{CNN}{convolutional neural network}
  \acro{FC}{fully connected}
  \acro{3x1DLS}[$3\x1$D-LS]{\ThreeDLS}
  \acro{2D-LS}{2-dimensional least-squares}
  \acro{CNN+LS}{neural network initialized 2-dimensional least-squares}
  \acro{LS}{least-squares}
\end{acronym}

\section{Introduction}\label{sec:intro}

    With usages ranging from tests of fundamental
    physics~\cite{Hiramoto2023,Aion2020},
    to exquisitely accurate atomic clocks~\cite{Aeppli2024} and potential platforms for
    universal quantum computers~\cite{evered2023,Giraldo2022}, ultracold atoms and molecules are the
    backbone of a wide range of modern physics experiments.
    In many ultracold atomic experiments, the primary observable of the
    quantum system is the spatial distribution of the atom cloud at a
    given time.  Absorption imaging of the cloud is often used to extract
    this spatial information. Accurate determination of the parameters which
    describe the spatial distribution of the cloud (such as the cloud location,
    shape, and number of atoms) is crucial to the success of many ultracold atom
    applications~\cite{Frye2021,Wilson2020,Squires2016}.

    Although absorption imaging is often the best way to extract desired data,
    it presents many challenges. Fluctuations in laser pointing and intensity
    as well as interferometric structures created by imperfections along the beam path
    can lead to noise and artifacts in the analysis of a final image.  Many
    efforts have been made over the years to address these problems and produce
    more accurate data~\cite{Pal23,Song2020,Niu2018}. 

    In recent years, deep learning has emerged as a powerful tool for image
    processing, offering significant improvements in speed and accuracy over
    traditional methods. \Acp{CNN} and other deep
    learning models have demonstrated remarkable success in tasks such as image
    classification, object detection, and image segmentation. Recently, \ac{ML}
    has been applied to experiments in various ways to improve the experimental
    process~\cite{Lode2021, Radovic2018, Baldi2016, Vajente2020}.
    In this paper, we present a deep learning-based method for extracting
    useful fit parameters from absorption image data.
    Deep learning has been previously applied to absorption
    imaging~\cite{Guo2021, Hofer2021, Ness2020}. Unlike these other methods, our
    method requires negligible user intervention, and is trained on simulated
    truth data.  We compare
    the accuracy and computation time of the deep learning models against
    \ac{LS} techniques.  We find the \ac{CNN} model achieves accuracy similar to
    \ac{LS}, while significantly reducing computation time.  We
    also compare the performance of models which take a single input atom image
    against models which use multiple input images (one atom and two background), finding that both models
    achieved similar accuracy.  The use of single image models enables
    single-exposure absorption imaging, which simplifies experiment
    design, and eases imaging hardware requirements.

\section{Background}\label{sec:background}

\subsection{Cold Atom Experiment}\label{sec:background:physics}

    In absorption imaging, an atom cloud is exposed to a collimated
    laser beam, which is resonant with a relevant atomic transition.  Photons
    are scattered by atoms in the beam path creating a shadow, which contains
    information regarding the spatial distribution of the atom cloud.
    This shadow is then imaged onto a camera.  We will denote this image as
    $\ssc{I}{atoms}$. To determine the fraction of light transmitted
    through the atom cloud, typically two additional images are taken:
    (1) an image of the beam
    in the absence of atoms, $\ssc{I}{bg}$, and (2) an image taken
    in the absence of both the atoms and the imaging beam, $\ssc{I}{dark}$.
    Light collected in $\ssc{I}{dark}$ is assumed to be non-resonant background
    light, which is common to both $\ssc{I}{atoms}$ and $\ssc{I}{bg}$.
    Consequently, $\ssc{I}{dark}$ is subtracted
    from both $\ssc{I}{atoms}$ and $\ssc{I}{bg}$.  The fraction of
    light transmitted through the atoms is calculated as
    \begin{equation}
      \label{eq:fraction}
      T(x,y) = \frac{\ssc{I}{atoms} - \ssc{I}{dark}}{\ssc{I}{bg} - \ssc{I}{dark}}
    \end{equation}

    Note that Eq.~\ref{eq:fraction} assumes that the only difference between
    $\ssc{I}{atoms}$ and $\ssc{I}{bg}$ arises from the presence
    of atoms in the imaging beam path.  In practice, the location, intensity
    and transverse mode of the imaging beam may change slightly between the
    $\ssc{I}{atoms}$ and $\ssc{I}{bg}$ images. Because transverse mode
    structure often arises due to diffractive effects, this noise is sensitive
    to changes in optical path length on the order of
    tens of nanometers.  We will refer to this noise as imaging beam noise. To
    help mitigate this noise, a typical experiment will take the $\ssc{I}{bg}$
    image as close-in-time as possible to the $\ssc{I}{atoms}$ image.

    The \ac{OD}, $\OD(x,y;\mathbf{p})$, is defined as the natural logarithm of $1/T(x,y)$.
    For a non-interacting ideal gas held in a harmonic potential, the
    \ac{OD} takes the form of a 2-dimensional Gaussian, which may be
    parameterized by $\mathbf{p} = \langle x_0, y_0, \sigma_x, \sigma_y, \rho, B, \theta \rangle$,
    where $(x_0, y_0)$ is the center position of the atom cloud,
    while $\sigma_x$ and $\sigma_y$ are the widths of a single standard deviation of the cloud
    measured along orthogonal principle axes rotated
    through an angle $\theta$ about the point $(x_0,y_0)$.
    To avoid overparameterization of the Gaussian,
    $\theta$ is restricted to $\left[-\pi/4, \pi/4\right)$, and we define $\sigma_x$ as the
    principle axis most aligned with the horizontal axis of the
    imaging system.  $B$ is an \ac{OD} offset which arises from imaging-beam
    noise, and $\rho$ is the peak \ac{OD}.

    In practice, the \ac{OD} of the atom cloud is calculated
    using images taken during an experiment, and the resulting \ac{OD}
    is fit to a 2-dimensional Gaussian.  The number of atoms, cloud
    size, and location can be calculated from these fit parameters.

\subsection{Neural Networks}\label{sec:background:nn}
\begin{figure*}[htb!]
  \centering
  \includegraphics[width=\textwidth]{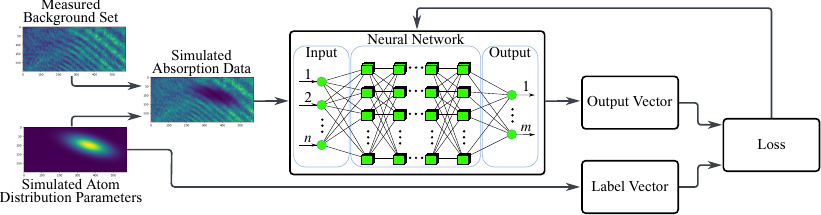}
  \caption{Overview of the deep learning method used to analyze absorption
  images of a cloud of cold atoms.  A representative set of experimentally
  measured background images is sampled and combined with a set of
  randomly parameterized atom-cloud distributions to generate simulated
  absorption data to train the neural network.
  }
  \label{fig:CNN_flowchart}
\end{figure*}

    Neural networks consist of a set of simulated neurons which are organized into
    layers.  The output of neurons from one layer are weighted together and
    fed into the neurons of the next layer.  The weights of each
    input/output pair are independent.  Neural networks learn to map inputs to
    outputs by adjusting these weights through an
    iterative process called training.

    During training, the model is provided with a set of
    input data, along with the corresponding known output values,
    referred to as truth values.
    A loss function compares the output
    of the model against known truth values.  Gradient descent
    techniques are used to minimize the loss with respect to the weights.

    In the work presented here, we use a \ac{CNN}.  In a \ac{CNN},
    the neurons are convolutional filters, which are squares of $n \times n$,
    usually ranging from $1 \times 1$ to $5 \times 5$ pixels.  The filter
    spans the image and preserves spatial relationships in image
    data since adjacent pixels are calculated together; this property
    makes \ac{CNN}s especially well-suited for image analysis.

\section{Theory of Operation}\label{sec:theory}

    While \ac{CNN}s are well-suited for predicting
    the parameters which define a Gaussian-shaped atom cloud from absorption
    image data, training a \ac{CNN} requires a large set of input
    data with known truth values.  This presents an obstacle as absorption image
    data is both time-consuming to collect and lacks known truth values.
    To address these challenges, this paper presents an option of training
    a \ac{CNN} on simulated absorption-image data.

    Towards this end, note that $\ssc{I}{atom}$ may be calculated
    according to Eq.~\ref{eq:fraction} as
    \begin{equation}
      \label{eq:atom_image}
      \ssc{I}{atoms} = T(x,y) \, (\ssc{I}{bg} - \ssc{I}{dark}) + \ssc{I}{dark} \, .
    \end{equation}
    The $\ssc{I}{bg}$ and $\ssc{I}{dark}$ images used to generate simulated
    $\ssc{I}{atoms}$ images are collected from the experiment.
    The parameters, $\mathbf{p}$, which define the \ac{OD},
    are chosen randomly from a set of values
    representative of the Gaussian atom clouds seen in
    the experiment.

    Simulated images used to train the model must be
    representative of actual images obtained by the experiment
    to ensure that the model adapts to real absorption image data.
    A challenge in creating simulated images is the accurate
    simulation of noise.  Simulated
    absorption images inherit noise from the real $\ssc{I}{bg}$ and
    $\ssc{I}{dark}$ images used to generate the simulated $\ssc{I}{atoms}$ image.
    No effort was made to add additional sources of noise to the simulated
    absorption images.  As we will show in  Sec.~\ref{sec:results}, the
    imaging noise inherited from the $\ssc{I}{bg}$ and
    $\ssc{I}{dark}$ images, was sufficient to allow the \ac{CNN}
    model to generalize to real absorption image data. To ensure the model is
    exposed to a representative sample of imaging beam noise, 835
    ($\ssc{I}{bg}$,$\ssc{I}{dark}$) pairs were used to train the model.

\section{Implementation}\label{sec:experiment}

    We studied the performance of several \ac{CNN} models,
    including MobiletNetV3 \cite{Howard20191314}, EfficientNet-B1, EfficientNet-B3 \cite{Tan2019},
    and RegNety-320 \cite{Radosavovic2020}. We found that the accuracy of these
    models are within the random variations which
    arise due to the stochastic nature of the training.
    In Sec.~\ref{sec:results}, we present results from MobiletNetV3
    as it achieves the lowest training and evaluation times.

    There are two versions of the model we will analyze here: ML-1 and ML-3.
    ML-1 is a \ac{CNN} trained to take only one image, $\ssc{I}{atoms}$, as an input
    while
    ML-3 takes three input images: $\ssc{I}{atoms}$, $\ssc{I}{bg}$ and
    $\ssc{I}{dark}$.  Both models output the parameter vector $\mathbf{p}$.
    Our loss
    function is defined as the mean of the square of the z-score normalized
    errors in $\mathbf{p}$.  The errors are z-score normalized
    according to the standard deviation of the distributions used to generate the
    simulated $\ssc{I}{atom}$ images.  This is done so that the model weights each
    fit parameter equally.  When generating simulated $\ssc{I}{atom}$ images, the
    parameters $\mathbf{p}$ are drawn from uniform distributions over some
    reasonable range which reflects the results seen by the real experiments.
    The range of values used to generate the simulated $\ssc{I}{atoms}$ images are
    summarized in Table~\ref{tab:param_dist}

    \begin{table}[h]
      \centering
      \begin{tabular}{@{}llll@{}}
        \toprule
        \textbf{Parameter} & \textbf{Minimum} & \textbf{Maximum} & \textbf{Units} \\
        \midrule
        $x_0$   & $0.1 \times {\rm W}$ & $0.9 \times {\rm W}$   & pixels\\
        $y_0$   & $0.1 \times {\rm W}$ & $0.9 \times {\rm H}$   & pixels \\
        $\sigma_x$   & $0$                  & $0.25 \times {\rm W}$  & pixels\\
        $\sigma_y$   & $0$                  & $0.25 \times {\rm H}$  & pixels\\
        $\rho$  & $0$                  & $3$                    & Optical Density\\
        $B$     & $-0.05$              & $0.05$                 & Optical Density\\
        $\theta$& $-0.1$               & $0.1$                  & radians\\
        \bottomrule
      \end{tabular}%
      \caption{
        Parameter ranges used in generating simulated $\ssc{I}{atoms}$ images.
        Note that W and H represent the width and height of the images,
        respectively.
        \label{tab:param_dist}
      }
    \end{table}

    The $\ssc{I}{bg}$ and $\ssc{I}{dark}$ images used to generate
    simulated $\ssc{I}{atoms}$ images are taken from
    ultra-cold atom experiments similar to those described in
    Ref.~\cite{Squires2016}.
    When training the ML-3 model, these images are organized
    into sets.  A $(\ssc{I}{bg}, \ssc{I}{dark})$ pair is
    collected during normal experiment runs, so that that the
    timing between the images is typical of normal operation.
    To correctly model imaging noise, the $\ssc{I}{bg}$ image used to generate
    the simulated $\ssc{I}{atoms}$ image must be different than the
    $\ssc{I}{bg}$ directly input into ML-3 while also being closely correlated
    in time. To this end, the $\ssc{I}{bg}$ used to generate $\ssc{I}{atoms}$ is
    generated using the $\ssc{I}{bg}$ image collected during a subsequent
    experimental run.

\section{Results}\label{sec:results}

There are two qualities of \ac{OD} image processing techniques
we will explore in the present analysis: accuracy and computation time.
These two qualities are generally in conflict, as improvements to
accuracy often come at the expense of increased computation time.

It is instructive to compare the performance of the \ac{CNN} model
against standard \ac{LS} techniques.
Towards this end, we will explore four \ac{OD} image analysis techniques,
labeled here as \acs{3x1DLS}, \acs{2D-LS}, ML-1, and ML-3.
The \ac{3x1DLS} algorithm iteratively performs 1-dimensional
\ac{LS} fits, alternating between horizontal and vertical
slices of the \ac{OD} data.  These slices pass through the cloud
center.  Earlier \ac{LS} fits provide subsequent fits with
improved estimates of the cloud center.  The \ac{3x1DLS} algorithm
does not account for cloud rotation.  The \ac{2D-LS} algorithm
is a 2-dimensional \ac{LS} fit, including rotation.  The \ac{2D-LS} method
applies the \ac{3x1DLS} algorithm to rotated horizontal and vertical axes
to provide initial estimates of the fit parameters.  The ML-1
algorithm uses the \ac{CNN} model trained on simulated $\ssc{I}{atoms}$
images only.  While ML-3 uses the \ac{CNN} model trained on three
images ($\ssc{I}{atoms}$, $\ssc{I}{bg}$, and
$\ssc{I}{dark}$). Note that the \ac{LS} fits and the \ac{ML} processing
presented in this section are performed on experimental absorption image data.

\begin{figure}
  \centering
  \includesvg[width=\columnwidth]{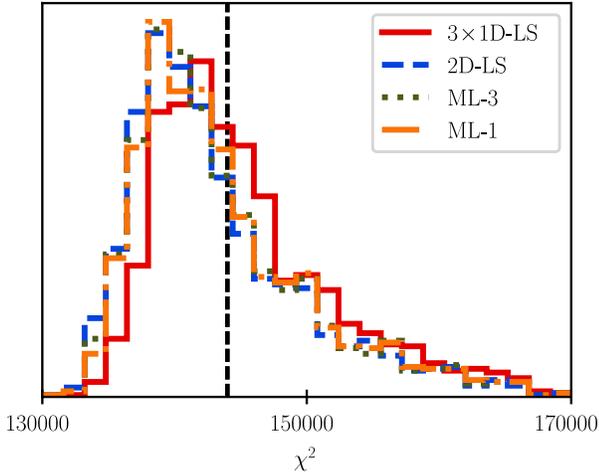}
  \caption{Distribution of $\chi^2$ values across 1392 absorption image fits.
           The vertical dashed line shows the degree of freedom.
  }
  \label{fig:chi_square_hist}
\end{figure}

We use the $\chi^2$ as a figure of merit to explore the
accuracy of the \ac{OD} analysis techniques.
The distribution of $\chi^2$ values, calculated across 1392 images, is shown
in Fig.~\ref{fig:chi_square_hist}, while the median $\chi^2$ values are
summarized in Table~\ref{tab:performance}.  Strikingly, both ML-1 and
ML-3 models achieve nearly the same $\chi^2$ as the \ac{2D-LS}, despite
the latter algorithm's explicit goal of minimizing $\chi^2$.  We also
observed that ML-1 achieves nearly the same chi-square as ML-3, suggesting that
the $\ssc{I}{dark}$ and $\ssc{I}{background}$ images provide almost no additional
information.

\begin{figure}
  \centering
  \includesvg[width=\columnwidth]{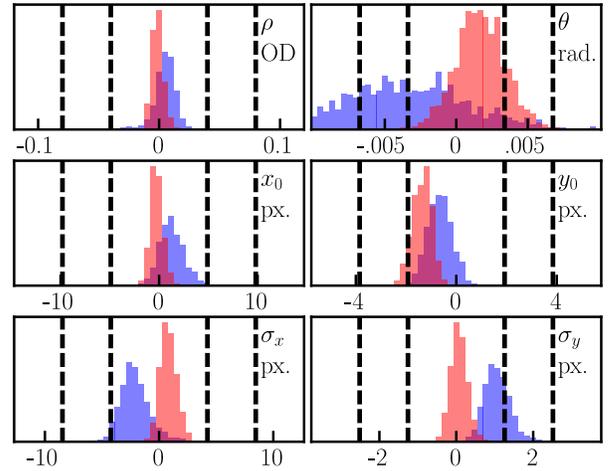}
  \caption{Distribution of parameter errors produced by ML-1 (blue) and ML-3 (red).
           The \ac{2D-LS} fit result is treated as truth.  Vertical lines represent $\pm 1$ and $\pm 2$
           standard deviations of the parameter observed over repeated experiment
           runs while experimental control parameters are held constant.  Units
           for the horizontal axis are shown in the upper right corner of each plot.
  }
  \label{fig:error_hist}
\end{figure}

The slightly larger $\chi^2$ achieved by both \ac{ML} models invites further investigation.
Towards this end, we compare each fit parameter output by the \ac{ML} models against
the results obtained by the \ac{2D-LS} method.  Treating the
\ac{2D-LS} results as the true values for each parameter, the distribution of the
errors produced by the \ac{ML} models are shown in Fig.~\ref{fig:error_hist}.  From this
analysis we see that the \ac{ML} models exhibit little systematic error.  While these
errors are outside the uncertainties in the \ac{LS} fit, our experiments have many
additional sources of random error, which can be measured by running the experiment
repeatedly while holding the experimental controls constant.  Comparing the
errors in the \ac{ML} models to the standard deviations of the parameters seen across
repeated experiment runs (with outlier fit parameters removed), shows that the
errors in the models are well within $\pm \sigma$ for all fit parameters with the
exception of $\theta$, which is still within $\pm 3\sigma$.  From this analysis
we conclude that the fitting error introduced by the \ac{CNN} model is small
compared to other sources of error in the experiment.

It is important to note that the training range for each parameter is
significantly larger than the ranges seen in typical experiments; this is to
increase the generality of the results. If the training parameters are more
closely tailored to specific experimental conditions, the accuracy of the model
is improved.

\begin{figure}
  \centering
  \includesvg[width=\columnwidth]{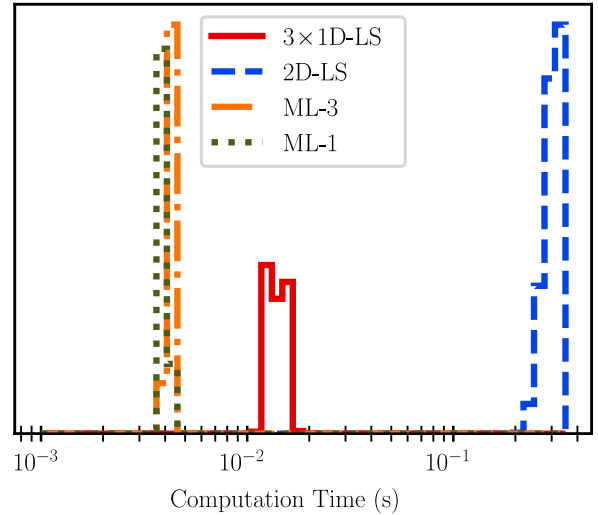}
  \caption{Distribution of fit computation times across 1392 absorption image fits.
          All computation times were recorded for a system with a Ryzen 7950X3D
          CPU and no GPU acceleration.
  }
  \label{fig:proc_time_hist}
\end{figure}

A distribution of the computation times is shown in Fig.~\ref{fig:proc_time_hist}
while median computation times are summarized in Table~\ref{tab:performance}.
All computation times were recorded on a system with a Ryzen 7950X3D
CPU and no GPU acceleration.
Unsurprisingly, we find the ML algorithm outperforms all other algorithms
in computation time, including the \ac{3x1DLS} algorithm, which is designed to
prioritize speed over accuracy.
Additionally, we find that ML-3 achieves computation times similar to ML-1.
The high accuracy and low computation
time of \ac{ML} models suggest that they are particularly well-suited for low-power
applications, where computational power is limited.

\begin{table}[h]
  \centering
  \begin{tabular}{l|c|r}
      \toprule
      \textbf{Method} & \textbf{Median $\chi^2$} & \textbf{Median Computation Time} \\
      \midrule
      \ac{3x1DLS} & 143831 &  $13 \ms$ \\
      \ac{2D-LS}  & 141651 & $328 \ms$ \\
      ML-1          & 141991 & $4 \ms$  \\
      ML-3          & 141896 &  $4 \ms$ \\
      \bottomrule
  \end{tabular}%
  \caption{Accuracy and computation time for various absorption image
  analysis techniques. All computation times were recorded for a
  system with a Ryzen 7950X3D CPU and no GPU acceleration.}\label{tab:performance}
\end{table}

In machine learning, fine tuning involves taking an already-trained model and
training it on more samples of a specific variety to improve the model's
accuracy in that domain. Changes in the
environment of the ultracold atom experiments can result in
modifications in the backgrounds used to train the model, which will
cause its accuracy to decrease over time.  This can be fixed by fine-tuning on
a small batch of atom images, which we found takes around five epochs, or one
minute of training on our specific machine, to converge to previous accuracy.
This is true even on a dataset of images taken months after the images that the
model was trained on. As such, we do not believe the variance in the cold atom
experiments over time to be an inhibiting factor to the usefulness of the model.
Unsurprisingly however, we found that, even with fine tuning, the model does not
generalize well from one experimental apparatus to another; in such cases a full
retraining is necessary.

\section{Conclusion}

We have demonstrated that \ac{CNN} models trained on simulated
absorption image data may be used to extract Gaussian fit
parameters from absorption image data.  We have compared
the \ac{CNN} models against \ac{LS} techniques and found
that they achieve a similar level of accuracy, while consuming
significantly fewer computational resources.  We explored the
efficacy of \ac{CNN} models which use a single atom image,
$\ssc{I}{atoms}$, and found that these models achieve nearly
the same accuracy as models which also rely on
background images ($\ssc{I}{bg}$, $\ssc{I}{dark}$).
The use of single-image models enable single-exposure absorption
imaging, which simplifies experiment design, and eases imaging
hardware requirements.

This technique could also be employed to more efficiently create a data mask,
for image analysis techniques that require them (e.g.\ principal component
analysis). In addition, our technique can be applied to any parameterizable
cloud shape, not just a Gaussian, as long as the shape can be appropriately
simulated for training.

The code used to train and evaluate the model, along with documentation,
can be found online:\\
\url{https://github.com/afrl-quantum/ml-imaging}.


\section*{Acknowledgments}
This work was partially funded by the Air Force Office of Scientific Research
under lab task 22RVCOR017.

\section*{Disclaimer}
The views expressed are those of the authors and do not necessarily reflect the
official policy or position of the Department of the Air Force, the Department
of the Defense, or the U.S. Government.

\bibliographystyle{apsrev4-2}
\bibliography{ml-imaging}

\begin{thebibliography}{21}%
\makeatletter
\providecommand \@ifxundefined [1]{%
 \@ifx{#1\undefined}
}%
\providecommand \@ifnum [1]{%
 \ifnum #1\expandafter \@firstoftwo
 \else \expandafter \@secondoftwo
 \fi
}%
\providecommand \@ifx [1]{%
 \ifx #1\expandafter \@firstoftwo
 \else \expandafter \@secondoftwo
 \fi
}%
\providecommand \natexlab [1]{#1}%
\providecommand \enquote  [1]{``#1''}%
\providecommand \bibnamefont  [1]{#1}%
\providecommand \bibfnamefont [1]{#1}%
\providecommand \citenamefont [1]{#1}%
\providecommand \href@noop [0]{\@secondoftwo}%
\providecommand \href [0]{\begingroup \@sanitize@url \@href}%
\providecommand \@href[1]{\@@startlink{#1}\@@href}%
\providecommand \@@href[1]{\endgroup#1\@@endlink}%
\providecommand \@sanitize@url [0]{\catcode `\\12\catcode `\$12\catcode
  `\&12\catcode `\#12\catcode `\^12\catcode `\_12\catcode `\%12\relax}%
\providecommand \@@startlink[1]{}%
\providecommand \@@endlink[0]{}%
\providecommand \url  [0]{\begingroup\@sanitize@url \@url }%
\providecommand \@url [1]{\endgroup\@href {#1}{\urlprefix }}%
\providecommand \urlprefix  [0]{URL }%
\providecommand \Eprint [0]{\href }%
\providecommand \doibase [0]{https://doi.org/}%
\providecommand \selectlanguage [0]{\@gobble}%
\providecommand \bibinfo  [0]{\@secondoftwo}%
\providecommand \bibfield  [0]{\@secondoftwo}%
\providecommand \translation [1]{[#1]}%
\providecommand \BibitemOpen [0]{}%
\providecommand \bibitemStop [0]{}%
\providecommand \bibitemNoStop [0]{.\EOS\space}%
\providecommand \EOS [0]{\spacefactor3000\relax}%
\providecommand \BibitemShut  [1]{\csname bibitem#1\endcsname}%
\let\auto@bib@innerbib\@empty
\bibitem [{\citenamefont {Hiramoto}\ \emph {et~al.}(2023)\citenamefont
  {Hiramoto} \emph {et~al.}}]{Hiramoto2023}%
  \BibitemOpen
  \bibfield  {author} {\bibinfo {author} {\bibfnamefont {A.}~\bibnamefont
  {Hiramoto}} \emph {et~al.},\ }\href
  {https://doi.org/10.1016/j.nima.2022.167513} {\bibfield  {journal} {\bibinfo
  {journal} {Nuclear Instruments and Methods in Physics Research Section A:
  Accelerators, Spectrometers, Detectors and Associated Equipment}\ }\textbf
  {\bibinfo {volume} {1045}},\ \bibinfo {pages} {167513} (\bibinfo {year}
  {2023})}\BibitemShut {NoStop}%
\bibitem [{\citenamefont {Badurina}\ \emph {et~al.}(2020)\citenamefont
  {Badurina} \emph {et~al.}}]{Aion2020}%
  \BibitemOpen
  \bibfield  {author} {\bibinfo {author} {\bibfnamefont {L.}~\bibnamefont
  {Badurina}} \emph {et~al.},\ }\href
  {https://doi.org/10.1088/1475-7516/2020/05/011} {\bibfield  {journal}
  {\bibinfo  {journal} {Journal of Cosmology and Astroparticle Physics}\
  }\textbf {\bibinfo {volume} {2020}}\bibinfo  {number} { (05)},\ \bibinfo
  {pages} {011}}\BibitemShut {NoStop}%
\bibitem [{\citenamefont {Aeppli}\ \emph {et~al.}(2024)\citenamefont {Aeppli},
  \citenamefont {Kim}, \citenamefont {Warfield}, \citenamefont {Safronova},\
  and\ \citenamefont {Ye}}]{Aeppli2024}%
  \BibitemOpen
\bibfield  {number} {  }\bibfield  {author} {\bibinfo {author} {\bibfnamefont
  {A.}~\bibnamefont {Aeppli}}, \bibinfo {author} {\bibfnamefont
  {K.}~\bibnamefont {Kim}}, \bibinfo {author} {\bibfnamefont {W.}~\bibnamefont
  {Warfield}}, \bibinfo {author} {\bibfnamefont {M.~S.}\ \bibnamefont
  {Safronova}},\ and\ \bibinfo {author} {\bibfnamefont {J.}~\bibnamefont
  {Ye}},\ }\href {https://doi.org/10.1103/PhysRevLett.133.023401} {\bibfield
  {journal} {\bibinfo  {journal} {Phys. Rev. Lett.}\ }\textbf {\bibinfo
  {volume} {133}},\ \bibinfo {pages} {023401} (\bibinfo {year}
  {2024})}\BibitemShut {NoStop}%
\bibitem [{\citenamefont {Evered}\ \emph {et~al.}(2023)\citenamefont {Evered}
  \emph {et~al.}}]{evered2023}%
  \BibitemOpen
  \bibfield  {author} {\bibinfo {author} {\bibfnamefont {S.~J.}\ \bibnamefont
  {Evered}} \emph {et~al.},\ }\href
  {https://www.nature.com/articles/s41586-023-06481-y} {\bibfield  {journal}
  {\bibinfo  {journal} {Nature}\ }\textbf {\bibinfo {volume} {622}},\ \bibinfo
  {pages} {268} (\bibinfo {year} {2023})}\BibitemShut {NoStop}%
\bibitem [{\citenamefont {Giraldo}\ \emph {et~al.}(2022)\citenamefont
  {Giraldo}, \citenamefont {Kumar}, \citenamefont {Wu}, \citenamefont {Du},\
  and\ \citenamefont {Weiss}}]{Giraldo2022}%
  \BibitemOpen
  \bibfield  {author} {\bibinfo {author} {\bibfnamefont {F.}~\bibnamefont
  {Giraldo}}, \bibinfo {author} {\bibfnamefont {A.}~\bibnamefont {Kumar}},
  \bibinfo {author} {\bibfnamefont {T.-Y.}\ \bibnamefont {Wu}}, \bibinfo
  {author} {\bibfnamefont {P.}~\bibnamefont {Du}},\ and\ \bibinfo {author}
  {\bibfnamefont {D.~S.}\ \bibnamefont {Weiss}},\ }\href
  {https://doi.org/10.1103/PhysRevA.106.032425} {\bibfield  {journal} {\bibinfo
   {journal} {Phys. Rev. A}\ }\textbf {\bibinfo {volume} {106}},\ \bibinfo
  {pages} {032425} (\bibinfo {year} {2022})}\BibitemShut {NoStop}%
\bibitem [{\citenamefont {Frye}\ \emph {et~al.}(2021)\citenamefont {Frye} \emph
  {et~al.}}]{Frye2021}%
  \BibitemOpen
  \bibfield  {author} {\bibinfo {author} {\bibfnamefont {K.}~\bibnamefont
  {Frye}} \emph {et~al.},\ }\href
  {https://epjquantumtechnology.springeropen.com/articles/10.1140/epjqt/s40507-020-00090-8}
  {\bibfield  {journal} {\bibinfo  {journal} {EPJ Quantum Technology}\ }\textbf
  {\bibinfo {volume} {8}},\ \bibinfo {pages} {1} (\bibinfo {year}
  {2021})}\BibitemShut {NoStop}%
\bibitem [{\citenamefont {Wilson}\ \emph {et~al.}(2020)\citenamefont {Wilson},
  \citenamefont {Malvania}, \citenamefont {Le}, \citenamefont {Zhang},
  \citenamefont {Rigol},\ and\ \citenamefont {Weiss}}]{Wilson2020}%
  \BibitemOpen
  \bibfield  {author} {\bibinfo {author} {\bibfnamefont {J.~M.}\ \bibnamefont
  {Wilson}}, \bibinfo {author} {\bibfnamefont {N.}~\bibnamefont {Malvania}},
  \bibinfo {author} {\bibfnamefont {Y.}~\bibnamefont {Le}}, \bibinfo {author}
  {\bibfnamefont {Y.}~\bibnamefont {Zhang}}, \bibinfo {author} {\bibfnamefont
  {M.}~\bibnamefont {Rigol}},\ and\ \bibinfo {author} {\bibfnamefont {D.~S.}\
  \bibnamefont {Weiss}},\ }\href {https://doi.org/10.1126/science.aaz0242}
  {\bibfield  {journal} {\bibinfo  {journal} {Science}\ }\textbf {\bibinfo
  {volume} {367}},\ \bibinfo {pages} {1461} (\bibinfo {year}
  {2020})}\BibitemShut {NoStop}%
\bibitem [{\citenamefont {Squires}\ \emph {et~al.}(2016)\citenamefont
  {Squires}, \citenamefont {Olson}, \citenamefont {Kasch}, \citenamefont
  {Stickney}, \citenamefont {Erickson}, \citenamefont {Crow}, \citenamefont
  {Carlson},\ and\ \citenamefont {Burke}}]{Squires2016}%
  \BibitemOpen
  \bibfield  {author} {\bibinfo {author} {\bibfnamefont {M.~B.}\ \bibnamefont
  {Squires}}, \bibinfo {author} {\bibfnamefont {S.~E.}\ \bibnamefont {Olson}},
  \bibinfo {author} {\bibfnamefont {B.}~\bibnamefont {Kasch}}, \bibinfo
  {author} {\bibfnamefont {J.~A.}\ \bibnamefont {Stickney}}, \bibinfo {author}
  {\bibfnamefont {C.~J.}\ \bibnamefont {Erickson}}, \bibinfo {author}
  {\bibfnamefont {J.~A.~R.}\ \bibnamefont {Crow}}, \bibinfo {author}
  {\bibfnamefont {E.~J.}\ \bibnamefont {Carlson}},\ and\ \bibinfo {author}
  {\bibfnamefont {J.~H.}\ \bibnamefont {Burke}},\ }\href
  {https://doi.org/10.1063/1.4971838} {\bibfield  {journal} {\bibinfo
  {journal} {Applied Physics Letters}\ }\textbf {\bibinfo {volume} {109}},\
  \bibinfo {pages} {264101} (\bibinfo {year} {2016})}\BibitemShut {NoStop}%
\bibitem [{\citenamefont {Pal}\ and\ \citenamefont {Chaudhuri}(2023)}]{Pal23}%
  \BibitemOpen
  \bibfield  {author} {\bibinfo {author} {\bibfnamefont {G.}~\bibnamefont
  {Pal}}\ and\ \bibinfo {author} {\bibfnamefont {S.}~\bibnamefont
  {Chaudhuri}},\ }\href {https://doi.org/10.1364/AO.503687} {\bibfield
  {journal} {\bibinfo  {journal} {Appl. Opt.}\ }\textbf {\bibinfo {volume}
  {62}},\ \bibinfo {pages} {8786} (\bibinfo {year} {2023})}\BibitemShut
  {NoStop}%
\bibitem [{\citenamefont {Song}\ \emph {et~al.}(2020)\citenamefont {Song},
  \citenamefont {He}, \citenamefont {Ren}, \citenamefont {Zhao}, \citenamefont
  {Lee},\ and\ \citenamefont {Jo}}]{Song2020}%
  \BibitemOpen
  \bibfield  {author} {\bibinfo {author} {\bibfnamefont {B.}~\bibnamefont
  {Song}}, \bibinfo {author} {\bibfnamefont {C.}~\bibnamefont {He}}, \bibinfo
  {author} {\bibfnamefont {Z.}~\bibnamefont {Ren}}, \bibinfo {author}
  {\bibfnamefont {E.}~\bibnamefont {Zhao}}, \bibinfo {author} {\bibfnamefont
  {J.}~\bibnamefont {Lee}},\ and\ \bibinfo {author} {\bibfnamefont {G.-B.}\
  \bibnamefont {Jo}},\ }\href
  {https://doi.org/10.1103/PhysRevApplied.14.034006} {\bibfield  {journal}
  {\bibinfo  {journal} {Phys. Rev. Appl.}\ }\textbf {\bibinfo {volume} {14}},\
  \bibinfo {pages} {034006} (\bibinfo {year} {2020})}\BibitemShut {NoStop}%
\bibitem [{\citenamefont {Niu}\ \emph {et~al.}(2018)\citenamefont {Niu},
  \citenamefont {Guo}, \citenamefont {Zhan}, \citenamefont {Chen},
  \citenamefont {Liu},\ and\ \citenamefont {Zhou}}]{Niu2018}%
  \BibitemOpen
  \bibfield  {author} {\bibinfo {author} {\bibfnamefont {L.}~\bibnamefont
  {Niu}}, \bibinfo {author} {\bibfnamefont {X.}~\bibnamefont {Guo}}, \bibinfo
  {author} {\bibfnamefont {Y.}~\bibnamefont {Zhan}}, \bibinfo {author}
  {\bibfnamefont {X.}~\bibnamefont {Chen}}, \bibinfo {author} {\bibfnamefont
  {W.~M.}\ \bibnamefont {Liu}},\ and\ \bibinfo {author} {\bibfnamefont
  {X.}~\bibnamefont {Zhou}},\ }\href {https://doi.org/10.1063/1.5040669}
  {\bibfield  {journal} {\bibinfo  {journal} {Appl. Phys. Lett.}\ }\textbf
  {\bibinfo {volume} {113}},\ \bibinfo {pages} {144103} (\bibinfo {year}
  {2018})}\BibitemShut {NoStop}%
\bibitem [{\citenamefont {Lode}\ \emph {et~al.}(2021)\citenamefont {Lode},
  \citenamefont {Lin}, \citenamefont {Büttner}, \citenamefont {Papariello},
  \citenamefont {Lévêque}, \citenamefont {Chitra}, \citenamefont {Tsatsos},
  \citenamefont {Jaksch},\ and\ \citenamefont {Molignini}}]{Lode2021}%
  \BibitemOpen
  \bibfield  {author} {\bibinfo {author} {\bibfnamefont {A.~U.}\ \bibnamefont
  {Lode}}, \bibinfo {author} {\bibfnamefont {R.}~\bibnamefont {Lin}}, \bibinfo
  {author} {\bibfnamefont {M.}~\bibnamefont {Büttner}}, \bibinfo {author}
  {\bibfnamefont {L.}~\bibnamefont {Papariello}}, \bibinfo {author}
  {\bibfnamefont {C.}~\bibnamefont {Lévêque}}, \bibinfo {author}
  {\bibfnamefont {R.}~\bibnamefont {Chitra}}, \bibinfo {author} {\bibfnamefont
  {M.~C.}\ \bibnamefont {Tsatsos}}, \bibinfo {author} {\bibfnamefont
  {D.}~\bibnamefont {Jaksch}},\ and\ \bibinfo {author} {\bibfnamefont
  {P.}~\bibnamefont {Molignini}},\ }\bibfield  {journal} {\bibinfo  {journal}
  {Phys. Rev. A}\ }\textbf {\bibinfo {volume} {104}},\ \href
  {https://doi.org/10.1103/PhysRevA.104.L041301} {10.1103/PhysRevA.104.L041301}
  (\bibinfo {year} {2021})\BibitemShut {NoStop}%
\bibitem [{\citenamefont {Radovic}\ \emph {et~al.}(2018)\citenamefont
  {Radovic}, \citenamefont {Williams}, \citenamefont {Rousseau}, \citenamefont
  {Kagan}, \citenamefont {Bonacorsi}, \citenamefont {Himmel}, \citenamefont
  {Aurisano}, \citenamefont {Terao},\ and\ \citenamefont
  {Wongjirad}}]{Radovic2018}%
  \BibitemOpen
  \bibfield  {author} {\bibinfo {author} {\bibfnamefont {A.}~\bibnamefont
  {Radovic}}, \bibinfo {author} {\bibfnamefont {M.}~\bibnamefont {Williams}},
  \bibinfo {author} {\bibfnamefont {D.}~\bibnamefont {Rousseau}}, \bibinfo
  {author} {\bibfnamefont {M.}~\bibnamefont {Kagan}}, \bibinfo {author}
  {\bibfnamefont {D.}~\bibnamefont {Bonacorsi}}, \bibinfo {author}
  {\bibfnamefont {A.}~\bibnamefont {Himmel}}, \bibinfo {author} {\bibfnamefont
  {A.}~\bibnamefont {Aurisano}}, \bibinfo {author} {\bibfnamefont
  {K.}~\bibnamefont {Terao}},\ and\ \bibinfo {author} {\bibfnamefont
  {T.}~\bibnamefont {Wongjirad}},\ }\href
  {https://doi.org/10.1038/s41586-018-0361-2} {\bibinfo {title} {Machine
  learning at the energy and intensity frontiers of particle physics}}
  (\bibinfo {year} {2018})\BibitemShut {NoStop}%
\bibitem [{\citenamefont {Baldi}\ \emph {et~al.}(2016)\citenamefont {Baldi},
  \citenamefont {Bauer}, \citenamefont {Eng}, \citenamefont {Sadowski},\ and\
  \citenamefont {Whiteson}}]{Baldi2016}%
  \BibitemOpen
  \bibfield  {author} {\bibinfo {author} {\bibfnamefont {P.}~\bibnamefont
  {Baldi}}, \bibinfo {author} {\bibfnamefont {K.}~\bibnamefont {Bauer}},
  \bibinfo {author} {\bibfnamefont {C.}~\bibnamefont {Eng}}, \bibinfo {author}
  {\bibfnamefont {P.}~\bibnamefont {Sadowski}},\ and\ \bibinfo {author}
  {\bibfnamefont {D.}~\bibnamefont {Whiteson}},\ }\bibfield  {journal}
  {\bibinfo  {journal} {Phys. Rev. D}\ }\textbf {\bibinfo {volume} {93}},\
  \href {https://doi.org/10.1103/PhysRevD.93.094034}
  {10.1103/PhysRevD.93.094034} (\bibinfo {year} {2016})\BibitemShut {NoStop}%
\bibitem [{\citenamefont {Vajente}\ \emph {et~al.}(2020)\citenamefont
  {Vajente}, \citenamefont {Huang}, \citenamefont {Isi}, \citenamefont
  {Driggers}, \citenamefont {Kissel}, \citenamefont {Szczepańczyk},\ and\
  \citenamefont {Vitale}}]{Vajente2020}%
  \BibitemOpen
  \bibfield  {author} {\bibinfo {author} {\bibfnamefont {G.}~\bibnamefont
  {Vajente}}, \bibinfo {author} {\bibfnamefont {Y.}~\bibnamefont {Huang}},
  \bibinfo {author} {\bibfnamefont {M.}~\bibnamefont {Isi}}, \bibinfo {author}
  {\bibfnamefont {J.~C.}\ \bibnamefont {Driggers}}, \bibinfo {author}
  {\bibfnamefont {J.~S.}\ \bibnamefont {Kissel}}, \bibinfo {author}
  {\bibfnamefont {M.~J.}\ \bibnamefont {Szczepańczyk}},\ and\ \bibinfo
  {author} {\bibfnamefont {S.}~\bibnamefont {Vitale}},\ }\bibfield  {journal}
  {\bibinfo  {journal} {Phys. Rev. D}\ }\textbf {\bibinfo {volume} {101}},\
  \href {https://doi.org/10.1103/PhysRevD.101.042003}
  {10.1103/PhysRevD.101.042003} (\bibinfo {year} {2020})\BibitemShut {NoStop}%
\bibitem [{\citenamefont {Guo}\ \emph {et~al.}(2021)\citenamefont {Guo},
  \citenamefont {Fritsch}, \citenamefont {Greenberg}, \citenamefont
  {Spielman},\ and\ \citenamefont {Zwolak}}]{Guo2021}%
  \BibitemOpen
  \bibfield  {author} {\bibinfo {author} {\bibfnamefont {S.}~\bibnamefont
  {Guo}}, \bibinfo {author} {\bibfnamefont {A.~R.}\ \bibnamefont {Fritsch}},
  \bibinfo {author} {\bibfnamefont {C.}~\bibnamefont {Greenberg}}, \bibinfo
  {author} {\bibfnamefont {I.~B.}\ \bibnamefont {Spielman}},\ and\ \bibinfo
  {author} {\bibfnamefont {J.~P.}\ \bibnamefont {Zwolak}},\ }\bibfield
  {journal} {\bibinfo  {journal} {Machine Learning: Science and Technology}\
  }\textbf {\bibinfo {volume} {2}},\ \href
  {https://doi.org/10.1088/2632-2153/abed1e} {10.1088/2632-2153/abed1e}
  (\bibinfo {year} {2021})\BibitemShut {NoStop}%
\bibitem [{\citenamefont {Hofer}\ \emph {et~al.}(2021)\citenamefont {Hofer},
  \citenamefont {Krstajić}, \citenamefont {Juhász}, \citenamefont
  {Marchant},\ and\ \citenamefont {Smith}}]{Hofer2021}%
  \BibitemOpen
  \bibfield  {author} {\bibinfo {author} {\bibfnamefont {L.~R.}\ \bibnamefont
  {Hofer}}, \bibinfo {author} {\bibfnamefont {M.}~\bibnamefont {Krstajić}},
  \bibinfo {author} {\bibfnamefont {P.}~\bibnamefont {Juhász}}, \bibinfo
  {author} {\bibfnamefont {A.~L.}\ \bibnamefont {Marchant}},\ and\ \bibinfo
  {author} {\bibfnamefont {R.~P.}\ \bibnamefont {Smith}},\ }\bibfield
  {journal} {\bibinfo  {journal} {Machine Learning: Science and Technology}\
  }\textbf {\bibinfo {volume} {2}},\ \href
  {https://doi.org/10.1088/2632-2153/abf5ee} {10.1088/2632-2153/abf5ee}
  (\bibinfo {year} {2021})\BibitemShut {NoStop}%
\bibitem [{\citenamefont {Ness}\ \emph {et~al.}(2020)\citenamefont {Ness},
  \citenamefont {Vainbaum}, \citenamefont {Shkedrov}, \citenamefont
  {Florshaim},\ and\ \citenamefont {Sagi}}]{Ness2020}%
  \BibitemOpen
  \bibfield  {author} {\bibinfo {author} {\bibfnamefont {G.}~\bibnamefont
  {Ness}}, \bibinfo {author} {\bibfnamefont {A.}~\bibnamefont {Vainbaum}},
  \bibinfo {author} {\bibfnamefont {C.}~\bibnamefont {Shkedrov}}, \bibinfo
  {author} {\bibfnamefont {Y.}~\bibnamefont {Florshaim}},\ and\ \bibinfo
  {author} {\bibfnamefont {Y.}~\bibnamefont {Sagi}},\ }\bibfield  {journal}
  {\bibinfo  {journal} {Phys. Rev. Appl.}\ }\textbf {\bibinfo {volume} {14}},\
  \href {https://doi.org/10.1103/PhysRevApplied.14.014011}
  {10.1103/PhysRevApplied.14.014011} (\bibinfo {year} {2020})\BibitemShut
  {NoStop}%
\bibitem [{\citenamefont {Howard}\ \emph {et~al.}(2019)\citenamefont {Howard}
  \emph {et~al.}}]{Howard20191314}%
  \BibitemOpen
  \bibfield  {author} {\bibinfo {author} {\bibfnamefont {A.}~\bibnamefont
  {Howard}} \emph {et~al.},\ }\href {https://doi.org/10.1109/ICCV.2019.00140}
  {\bibfield  {journal} {\bibinfo  {journal} {Proceedings of the IEEE
  International Conference on Computer Vision}\ ,\ \bibinfo {pages} {1314}}
  (\bibinfo {year} {2019})}\BibitemShut {NoStop}%
\bibitem [{\citenamefont {Tan}\ and\ \citenamefont {Le}(2019)}]{Tan2019}%
  \BibitemOpen
  \bibfield  {author} {\bibinfo {author} {\bibfnamefont {M.}~\bibnamefont
  {Tan}}\ and\ \bibinfo {author} {\bibfnamefont {Q.~V.}\ \bibnamefont {Le}},\
  }\href@noop {} {\bibfield  {journal} {\bibinfo  {journal} {International
  Conference on Machine Learning, 2019}\ } (\bibinfo {year} {2019})},\ \Eprint
  {https://arxiv.org/abs/0902.0885} {arXiv:0902.0885 [cs]} \BibitemShut
  {NoStop}%
\bibitem [{\citenamefont {Radosavovic}\ \emph {et~al.}(2020)\citenamefont
  {Radosavovic}, \citenamefont {Kosaraju}, \citenamefont {Girshick},
  \citenamefont {He},\ and\ \citenamefont {Dollár}}]{Radosavovic2020}%
  \BibitemOpen
  \bibfield  {author} {\bibinfo {author} {\bibfnamefont {I.}~\bibnamefont
  {Radosavovic}}, \bibinfo {author} {\bibfnamefont {R.~P.}\ \bibnamefont
  {Kosaraju}}, \bibinfo {author} {\bibfnamefont {R.}~\bibnamefont {Girshick}},
  \bibinfo {author} {\bibfnamefont {K.}~\bibnamefont {He}},\ and\ \bibinfo
  {author} {\bibfnamefont {P.}~\bibnamefont {Dollár}},\ }\href@noop {} {\
  (\bibinfo {year} {2020})},\ \Eprint {https://arxiv.org/abs/2003.13678}
  {arXiv:2003.13678 [cs]} \BibitemShut {NoStop}%
\end{thebibliography}%

\end{document}